# Models for the nonsingular transition of an evaporating black hole into a white hole

James M. Bardeen

*Physics Department, Box 1560, University of Washington*
*Seattle, Washington 98195-1560. USA*
*bardeen@uw.edu*

**Abstract**

There have been a number of suggestions that the $r = 0$ singularity of a spherically symmetric (uncharged) evaporating black hole can be circumvented by a quantum transition to a white hole, which eventually releases all trapped quantum information, consistent with overall unitary evolution of the quantum fields. Some of these scenarios rely on loop quantum gravity to impose a minimum area of two-spheres, but are rather vague on how to deal with black hole evaporation, particularly its endpoint. In this paper I present a rather complete toy model for the evolution of the geometry and the effective stress-energy tensor derived from the geometry via the classical Einstein equations. Modifications of the Schwarzschild geometry once the formation of the black hole is complete are very small where the spacetime curvature is small compared with the Planck scale, and the curvature never becomes super-Planckian. The evolution of the white hole is roughly the time reverse of the formation and evaporation of the black hole. The mass of the white hole increases as it gradually emits the negative energy that flowed into the black hole during its evaporation until the matter and radiation that collapsed to form the black hole emerges and the white hole disappears. I consider the compatibility of the model with some of the quantum energy conditions proposed in the literature, and the implications for the interpretation of black hole entropy.

## I. INTRODUCTION

The discovery of Hawking radiation from black holes[1] over 40 years ago led to the assertion[2] of a fundamental breakdown of predictability in the evolution of quantum fields following gravitational collapse to form a black hole. Quantum information is lost inside the event horizon of the black hole and apparently cannot be recovered, leading to a breakdown of unitarity. Of course, it is not at all surprising that separation of a quantum system into two parts (the exterior and interior of the black hole) results in each part considered separately being in a mixed state when the complete system is in a pure state, as argued strongly by Unruh and Wald[3]. However, complete evaporation of the black hole without release of the trapped quantum information does raise serious issues, particularly in the light of the AdS/CFT conjecture[4], in which gravity in the bulk is supposed to be dual to a manifestly unitary conformal field theory on the AdS boundary.

The thermodynamic entropy of a black hole interacting with its surroundings[5] is identified with the Bekenstein-Hawking entropy proportional to the area of the event horizon. In units with $G = c = 1$, $S_{BH} = A/(4\hbar)$. Normally the thermodynamic entropy of a quantum system is identified with its total number of quantum degrees of freedom, which in turn is the maximum possible value of the entanglement (von Neumann) entropy $S_{vN}$. If the Hawking radiation is entangled with degrees of freedom inside the black hole, as in the standard semi-classical theory of Hawking radiation, Page[6] has shown that $S_{vN}$ becomes equal to $S_{BH}$ at the Page time, when the black hole has lost only about one half of it initial mass. If the black hole continues emitting Hawking radiation after the Page time, as one would expect for any black hole with a mass much greater than the Planck mass $m_p$, either $S_{vN} > S_{BH}$ or the late Hawking radiation must be entangled with the early Hawking radiation. If the latter, by the monogamy of entanglement the late Hawking radiation cannot be entangled with Hawking "partners" inside the black hole horizon, resulting in a "firewall" of highly excited quanta propagating on or just inside the black hole horizon[7].

Controversy over these issues has raged right up to the present time, with no widely accepted resolution. See reviews by Marolf[8] and Polchinski[9]. A big part of the problem is the lack of a widely accepted theory of quantum gravity. Naively, for very large black holes the semi-classical theory of quantum fluctuations propagating on a classical geometry should be an excellent approximation. Tidal accelerations at the horizon of a very large astrophysical black hole are no larger than those in laboratories on the Earth, where quantum field theory has been tested with exquisite precision. I have argued at length elsewhere[10] that the semi-classical physics in the vicinity of the horizon of a large black hole precludes any substantial storage of quantum information on or near the horizon, and that almost all of the quantum information entangled with the Hawking radiation ends up in the deep interior of the black hole.

However, that does not mean the quantum information is irretrievably swallowed up by a singularity. The classical singularity theorems rely on energy conditions that are violated in quantum field theory. Various more or less ad hoc nonsingular black hole models, some inspired by loop quantum gravity (LQG)[11], have been proposed. One possibility is that quantum backreaction simply stops collapse short of a singularity, which requires an inner trapping horizon. If the inner and outer trapping horizons eventually merge and disappear, the quantum information can escape, as suggested by Hayward[12]. More or less similar models have been proposed by Hossenfelder, et al[13], Rovelli and Vidotto[14], Frolov[15], De Lorenzo, et al[16], and Bardeen[17]. Release of quantum information by the Page time requires a large quantum backreaction in regions of low curvature. The negative surface gravity of the inner trapping horizon raises questions about the viability of these models.

An interesting alternative is the conversion of the black hole into a white hole, as discussed in general terms by Modesto[18] and by Ashtekar and Bojowald[19]. More explicit models are in references [20,,21,22,23,24]. In some of these there is a Cauchy horizon to the future of the black hole interior, which does not resolve the unitarity issues.  What is required is a nonsingular quantum transition from the black holed to a white hole jn a spacetime with the causal structure of Minkowski spacetime.  The trapped quantum information escapes from the white hole and propagates out to future null infinity.  Models that invoke quantum tunneling directly from a large black hole directly to a large white hole, such as that of Haggard and Rovelli[22], I find less convincing than those with a smooth transition of the geometry at the Planck scale.  Bianchi, et al[23] (BCDHR) proposed an ad hoc metric, reviewed briefly in Part II, in which the circumferential radius has a nonzero minimum, without a singularity, at the transition from the black hole to the white hole.  Ashtekar, et al[24] (AOS) based such a metric more explicitly on LQG, but in neither case does the metric account for the global geometry of an *evaporating* black hole.  Is the evolution of the white hole roughly the time reverse of the evolution of the black hole, or does instability of the white hole horizon, as claimed by De Lorenzo and Perez[25], require a very short lifetime for the white hole?  BCDHR argued for a very small and very long-lived Planck-scale white hole.  There are also conflicting claims about the time scale of a tunneling process[26].

In Part III I present an improved, more complete model for the transition from a evaporating black hole to a white hole, in which the geometry varies smoothly across the transition, full account is taken of the Hawking radiation and its effect on the black hole geometry, and the effective stress-energy tensor calculated from the geometry with the classical Einstein equations extrapolates to the form of the semi-classical stress-energy tensor (SCSET) outside the black hole horizon.  For an external observer, the creation of the white hole takes place only after the black hole has evaporated down to the Planck scale, and the white hole evolution is roughly the time-reverse of the black hole evaporation.  The model is consistent with some of the semi-classical quantum energy conditions and has what is classically a stable white hole horizon, as discussed in Part IV.  However, the model requires that the white hole emit negative energy radiation propagating out to future null infinity.  Arguments for why this may be acceptable, or even necessary, and the implications for the interpretation of black hole entropy and the information problem, are considered in Part V.

## II. THE BCDHR GEOMETRY

BCDHR make a simple ansatz for an effective spherically symmetric metric that smoothly transitions from the black hole to the white hole.  The basic idea, inspired by LQG, is that the circumferential radius $r$ in the quantum geometry has a minimum value $l$.  This is implemented by setting $r \equiv \tau^2 + l$, with $\tau < 0$ on the black hole side and $\tau > 0$ on the white hole side. Relabeling the Schwarzschild $t$ as $x$, in view of its role as a spatial coordinate, the metric they propose is

$$ds^2 = -\frac{4(\tau^2 + l)}{2M - \tau^2} d\tau^2 + \frac{2M - \tau^2}{\tau^2 + l} dx^2 + (\tau^2 + l)^2 d\Omega^2. \tag{2.1}$$

There is a coordinate singularity at $\tau^2 = 2M$, but the curvature is finite everywhere. The metric is approximately Schwarzschild for $\tau^2 \gg l$, and $M$ is approximately the gravitational mass of the system. They assume that $l \sim \left(Mm_p^2\right)^{1/3}$, so that $\tau^2 \sim l$ is when the spacetime curvature approaches the Planck scale. This is considered the onset of tunneling from the black hole to the white hole. The parameter $M$ is treated as a constant, though this can only be an approximation valid locally while $M \gg m_p$ for a slowly evaporating black hole.

The coordinate singularity at the black hole horizon can be eliminated by changing, for $\tau \leq 0$, to an advanced null coordinate $v$, constant on ingoing radial null geodesics, with

$$v = x + 2\int_0^\tau \frac{\left(\tau^2 + l\right)^{3/2}}{2M - \tau^2} d\tau. \tag{2.2}$$

For $\tau \geq 0$ the retarded null coordinate $u$ removes the singularity at the white hole horizon

$$u = -x + 2\int_0^\tau \frac{\left(\tau^2 + l\right)^{3/2}}{2M - \tau^2} d\tau, \tag{2.3}$$

with $u = -v$ at the $\tau = 0$ transition from the black hole to the white hole.

BCDHR argue that the white hole should have a small Planck-scale mass and emit the quantum information trapped by the black hole over a time long compared with the black hole evaporation time. However, the parameters $M$ and $l$ decrease by enormous factors during the black hole evaporation, and Eq. (2.1) suggests that at an early stage of evaporation the interior of the black hole and the interior of the white hole should have the same large mass.

## III. MODELING AN EVAPORATING BLACK HOLE

In constructing a model for the evolution of the geometry of an evaporating black hole and the transition to a white hole, assuming spherical symmetry, it is highly advantageous to work in Eddington-Finkelstein coordinates. The advanced version of the metric is regular on the black hole horizon and the retarded version is regular on the white hole horizon. Furthermore, as pointed out by Bardeen[27], the Einstein equations for a general spherically symmetric metric are remarkably simple. The general advanced version of the metric, using the circumferential radius $r$ as a coordinate, is

$$ds^2 = -\left(1 - 2m/r\right)e^{2\psi_v}dv^2 + 2e^{\psi_v}dvdr + r^2d\Omega^2, \tag{3.1}$$

and the retarded version is

$$ds^2 = -\left(1 - 2m/r\right)e^{2\psi_u}du^2 - 2e^{\psi_u}dudr + r^2d\Omega^2. \tag{3.2}$$

The Misner-Sharp mass function $m$ has coordinate-independent defnition $\nabla_\alpha r \nabla^\alpha r = 1 - 2m/r$. The Einstein equations relating $m(v,r)$ and $\psi_v(v,r)$ to the stress-energy tensor are

$$-4\pi T^v_v = \frac{1}{r^2}\left(\frac{\partial m}{\partial r}\right)_v, \quad 4\pi T^r_v = \frac{1}{r^2}\left(\frac{\partial m}{\partial v}\right)_r, \quad 4\pi T^v_r = \frac{1}{r}\left(\frac{\partial e^{-\psi_v}}{\partial r}\right)_v, \qquad (3.3)$$

and in the retarded coordinates $m(u,r)$ and $\psi_u(u,r)$ satisfy

$$-4\pi T^u_u = \frac{1}{r^2}\left(\frac{\partial m}{\partial r}\right)_u, \quad 4\pi T^r_u = \frac{1}{r^2}\left(\frac{\partial m}{\partial u}\right)_r, \quad 4\pi T^u_r = \frac{1}{r}\left(\frac{\partial e^{-\psi_u}}{\partial r}\right)_u. \qquad (3.4)$$

Some models for black hole interiors suggested on the basis of loop quantum gravity postulate a mass function similar to

$$m = \frac{Mr^3}{r^3 + 2Ma^2}. \qquad (3.5)$$

In particular, Hayward[12] based a non-singular model of an evaporating black hole on Eq. (3.5) with $M = M(v)$ in the black hole interior, $e^{\psi_v} = 1$, and $a$ a constant the order of the Planck length. The Hayward model geometry is regular at $r = 0$ and curvature invariants, such as the square of the Riemann tensor, never become larger than Planckian in magnitude. However a non-singular transition to a white hole requires a minimum value of $r$, as in BCDHR and AOS.

Where I differ from BCDHR and AOS is that my minimum radius, which really should be correspond to a minimum two-sphere area, has a fixed Planck scale value $a$, independent of the mass of the black hole. Instead of using $r$ as a coordinate, I define a coordinate $z$ such that

$$r^2 = z^2 + a^2. \qquad (3.6)$$

The coordinate $z$ is by definition negative in the black hole and positive in the white hole, with a smooth transition at $z = 0$. With $z$ instead of $r$ as a coordinate, the advanced form of the metric given in Eq. (3.1) becomes

$$ds^2 = -e^{2\bar{\psi}_v} g^{zz} dv^2 - 2e^{\bar{\psi}_v} dv\,dz + r^2 d\Omega^2, \quad g^{zz} = \frac{r^2}{z^2}\left(1 - \frac{2m}{r}\right), \quad e^{\bar{\psi}_v} = \frac{|z|}{r} e^{\psi_v}. \qquad (3.7)$$

The retarded form corresponding to Eq. (3.2) becomes

$$ds^2 = -e^{2\bar{\psi}_u} g^{zz} du^2 - 2e^{\bar{\psi}_u} du\,dz + r^2 d\Omega^2, \quad g^{zz} = \frac{r^2}{z^2}\left(1 - \frac{2m}{r}\right), \quad e^{\bar{\psi}_u} = \frac{|z|}{r} e^{\psi_u}. \qquad (3.8)$$

Eq. (3.6) is equivalent to the expression for $g_{\theta\theta}$ derived from LQG in AOS,

$$r^2 = \left(2Me^T\right)^2 + \frac{1}{4}\frac{(\gamma L_0 \delta_c)^2 M^2}{(2Me^T)^2}. \qquad (3.9)$$

Here $M$ is the black hole mass, while $\gamma L_0$ and $\delta_c$ are LQG parameters and $T$ is a coordinate ranging from $+\infty$ far outside the black hole to $-\infty$ far outside the white hole. The minimum radius and coordinate $z$ of Eq. (3.6) corresponding to Eq. (3.9) are

$$a = (\gamma L_0 \delta_c M)^{1/2}, \quad z = -2Me^T + \frac{\gamma L_0 \delta_c}{4e^T}. \qquad (3.10)$$

In AOS the LQG parameters are chosen to give a minimum radius similar to that in BCDHR, with

$$L_0 \delta_c = \frac{1}{2}\left(\frac{\gamma \Delta^2}{4\pi^2 M}\right)^{1/3}, \tag{3.11}$$

where $\Delta$ is the fundamental "area gap" parameter of LQG, of order $\hbar$. The other LQG parameters have more arbitrary values. If, instead of Eq. (3.11), $\gamma L_0 \delta_c \simeq \Delta / M$, then $a \simeq \sqrt{\Delta}$ and is independent of the mass. The traditional choice is motivated by the desire to ensure that the spacetime curvature never becomes super-Planckian, but my model also satisfies this constraint for certain ranges of the parameters in my ansatz for the metric functions.

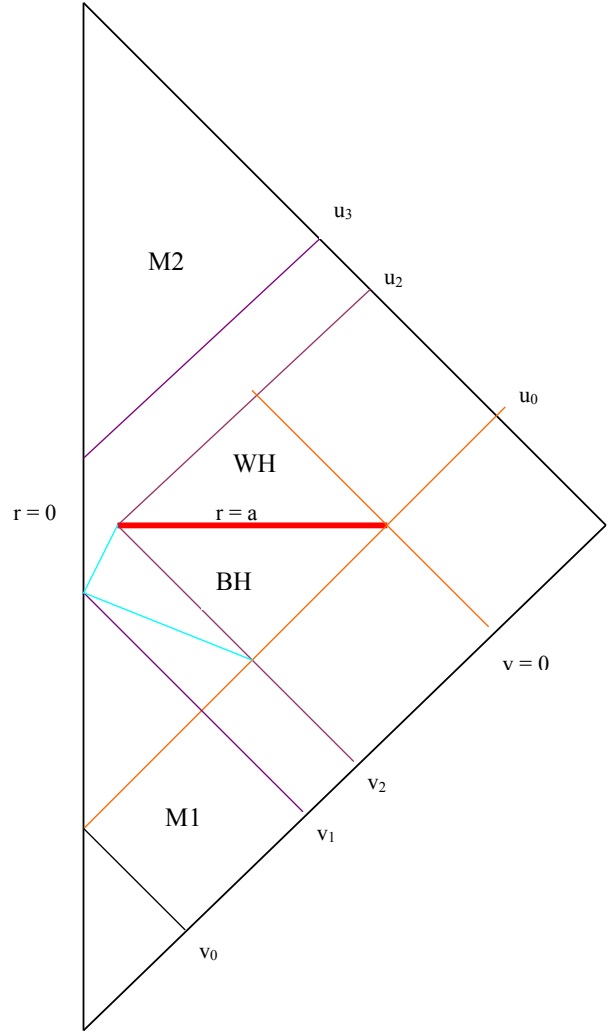

Figure 1. A Penrose diagram showing a thick null shell coming in from past null infinity over a range of advanced times $v_1 < v < v_2$. The black hole horizon is the portion of the $u = u_0$ null hypersurface extending from $v = v_0$ to $v = 0$. The transition to the white hole is at the $r = a$ spacelike hypersurface. See the text for further details.

The causal relationships in my model are illustrated in the Penrose diagram of Fig. 1. The black hole is formed by an influx of radiation along radial null geodesics in a "thick" null shell of mass $M_0$ between advanced times $v_1$ and $v_2$. An infinitesimally thin shell is not physically realistic when resolving geometry at close to the Planck scale. The black hole starts evaporating at $v = v_2$, with the dynamic horizon (*not* an event horizon) a null hypersurface at retarded time $u = u_0$, whose radius slowly decreases, ending by definition at $v = 0$, where $g^{zz} = 1$ at $r = a$ and $2M = 2M_{min}$ is of order $a$. There should be no trapped surfaces for any $z < 0$ at $v = 0$, in order to avoid any ambiguity in locating the end of the black hole and the creation of the white hole at $u = u_0$, $v = 0$. The horizon of the white hole is at $v = 0$ for $u > u_0$. The transition to the white hole is on the $z = 0$ spacelike hypersurface, where $g^{zz} < 0$ and $r = a$, indicated by the horizontal red line. There are trapped surfaces inside the region marked "BH" and anti-trapped surfaces inside the region marked "WH".

I make no attempt to explicitly model the dynamics of the radiation and evolution of the geometry in the interior of the shell, except that inside of the inner edge of the shell the geometry should be Minkowski (region M1). When the inner edge reaches $r = 0$, quantum backreaction must generate an inner trapping horizon if a singularity is to be avoided. This inner trapping horizon is a timelike hypersurface and should connect with the inner edge of the $r = a$ spacelike hypersurface at the outer surface of the shell, as indicated by the upper blue line. Instability of this inner trapping horizon is not a problem, since it only lasts for of order one e-folding of the blueshift. The energy in the shell is free to start flowing out to the future of the $u = u_2$ null hypersurface. If the outflow is complete by $u = u_3$, there is another Minkowski region M2 to the future of $u_3$. The other blue line represents an outer trapping horizon in the interior of the shell, a spacelike hypersurface that joins with the outer trapping horizon of the evaporating black hole, a timelike hypersurface just slightly outside the black hole horizon.

The scaling of Fig. 1 is extremely uneven. The advanced time $-v_2$ over which the black hole evaporates is $\sim M_0^3 / \hbar$, enormously greater than the range of advanced time $\sim M_0$ over which the black hole forms, and the bounce of the collapsing shell presumably takes place. Also, the Hawking radiation reaches future null infinity over what appears as an infinitesimal range of retarded time in the diagram, but which is actually comparable to $-v_2$ as measured by a distant observers.

My ansatz for the metric functions $g^{zz}$ and $e^{\psi_v}$ in the region outside the shell is is in the spirit of Eq. (3.5), but with the added flexibility to match the form of the SCSET in the exterior of the black hole suggested by numerical calculations for spin 0 and spin 1 fields[10]. Unfortunately, the spin 2 (graviton) contribution to the SCSET has not been calculated and presumably dominates, since the spin 2 trace anomaly is more than 10 times the spin1 trace anomaly in magnitude. Both metric functions are regular functions of $z$ at $z = 0$, implying $1 - 2m/r = 0$ there.

The expression for $g^{zz}$ is

$$g^{zz} = 1 - \frac{2Mr^2 + \alpha a^2 r}{r^3 + \beta a^2 r + 2\gamma M a^2}. \tag{3.12}$$

For a large ($M \gg a$) slowly evaporating black hole, the metric is Schwarzschild at $r \gg (Ma^2)^{1/3}$ and $M$ is the black hole mass. At $r \gg 2M$, where the Hawking radiation propagates on outward radial null geodesics, $M$ should be a function of retarded time. Calculations of the semi-classical stress-energy tensor (SCSET) show[10] that close to the black hole horizon there is an inflow of negative energy, decreasing the black hole mass in accord with the energy lost in Hawking radiation. Just how this negative energy propagates deep inside the black hole is somewhat uncertain, but I will assume that at $r \ll 2M$ $M$ is a function of advanced time. That is, $M = M(t')$, where $t'$ interpolates between an advanced time deep inside the black hole and a retarded time well outside the black hole horizon. The parameters $\alpha$, $\beta$, and $\gamma$ could vary with $t'$, but for simplicity I will take them to be constants.

Specifically, for $r > 2M$ I define the time $t'$ in the advanced Eddington-Finkelstein space-time coordinates of Eq. (3.1) as

$$t' = v - 2r - 4M(t')\ln[r/2M(t')], \tag{3.13}$$

so $t'$ at large $r$ is a Schwarzschild retarded time, but is finite, $t' = v - 4M$, at $r = 2M$. Inside the black hole, at $r < 2M$, I define

$$t' = v - r^4/(2M)^3 - 2M, \tag{3.14}$$

in accord with $C^1$ continuity. The derivative $dM/dt' = -L_H$, where $L_H$ is the Hawking luminosity. Then for $r > 2M$,

$$\left(\frac{\partial M}{\partial v}\right)_r = \frac{-L_H}{1 + 4L_H[1 + \ln(2M/r)]}, \quad \left(\frac{\partial M}{\partial r}\right)_v = \frac{-2 - 2(2M/r)}{1 + 4L_H[1 + \ln(2M/r)]}. \tag{3.15}$$

For $r < 2M$,

$$\left(\frac{\partial M}{\partial v}\right)_r = \frac{-L_H}{1 + L_H[6(r/2M)^4 - 2]}, \quad \left(\frac{\partial M}{\partial r}\right)_v = \frac{-4(r/2M)^3}{1 + L_H[6(r/2M)^4 - 2]}. \tag{3.16}$$

There is $C^0$ continuity in these derivatives at $r = 2M$. The denominators in Eqs. (3.15) and (3.16) are quite close to 1, since the semi-classical estimates are that $L_H$ should not be larger than about $0.001(a/2M)^2$, and I assume $L_H$ goes smoothly to zero at the end of the black hole evaporation.

The metric function $\bar{\psi}_v$ is related to $T_r^v$ by the last of Eqs. (3.3). Well outside the black hole horizon, in retarded coordinates, the only component of the stress-energy tensor falling as slowly as $r^{-2}$ is $T_u^r \cong -L_H/(4\pi r^2)$. For a large black hole the geometry is Schwarzschild to a good approximation and transforming to the advanced time coordinate gives

$$T_r^v \cong -4T_u^v \cong 4L_H / (4\pi r^2) + O(r^{-3}). \tag{3.17}$$

My ansatz for $e^{-\bar{\psi}_v}$ is consistent with this and introduces three additional parameters $\delta$, $\varepsilon$, and $\phi$, for simplicity assumed to be constants. For $r > 2M$

$$e^{-\bar{\psi}_v} = \left\{1 + 4L_H\left[1 + \ln\left(\frac{2M}{r}\right)\right]\right\}\left[1 + \delta\frac{a^2}{2Mr} + \varepsilon\frac{a^2}{r^2} + \phi\frac{2Ma^2}{r^3}\right]. \tag{3.18}$$

The first factor is just $(\partial t'/\partial v)_r^{-1}$, the denominator in Eqs. (3.15), and for $r < 2M$ is replaced by the denominator in Eqs. (3.16). In evaluating the radial derivatives of $\bar{\psi}_v$ I will, for simplicity, ignore the small derivatives of $M$ and $L_H$.

At the $z = 0$ transition to the white hole I switch to retarded Eddington-Finkelstein coordinates and assume that the negative energy associated with the Hawking "partners" flowing into the black hole flows out at constant retarded time $u$. This means $t'$ should be constant at constant $u$, and the white hole retarded time can be defined such that $u = -t'$. Then at $z = 0$,

$$u = -t', = -v + a^4/(2M)^3 + 2M, \tag{3.19}$$

and for $z > 0$, instead of Eqs. (3.16),

$$(\partial M/\partial u)_r = -dM/dt' = +L_H, \quad (\partial M/\partial r)_u = 0. \tag{3.20}$$

All the parameters are assumed to be continuous across the transition, which means $g^{zz}$ is continuous, but because $(\partial t'/\partial u)_z \neq -(\partial t'/\partial v)_z$, $e^{-\bar{\psi}_u} \neq e^{-\bar{\psi}_v}$ at $z = 0$. From the definition of $u$ in Eq. (3.19) the first factor in $e^{-\bar{\psi}_u}$ is just 1. A transition to a white hole requires that the $z = 0$ hypersurface be spacelike, $g^{zz} < 0$. The 2-surfaces with $g^{zz} < 0$ on the white hole side are anti-trapped surfaces, as opposed to the trapped surfaces on the black hole side.

At the endpoint of black hole evaporation, $v = 0$, $g^{zz} = 0$ at $r = a$ and $2M(t_0')/a \equiv (2M/a)_{\min} > 0$ is of $O(1)$. Eq. (3.19) then gives the value $u = u_0$ on the black hole horizon. The parameters should be chosen so that there are no trapped surfaces at $v = 0$ for any $z < 0$. This requires $(\partial g^{zz}/\partial r)_v > 0$ at $r = a$. From Eq. (3.12) $g^{zz} = 0$ implies

$$\left(\frac{2M}{a}\right)_{\min} = \frac{1 + \beta - \alpha}{1 - \gamma} > 0, \tag{3.21}$$

and $(\partial g^{zz}/\partial r)_{r=a} > 0$ if

$$1 - \beta + \alpha - \gamma(3 + \beta - \alpha) > 0. \tag{3.22}$$

It seems reasonable to require $0 < \gamma < 1$, and $\gamma \geq 0$ is necessary to avoid a singular $g^{zz}$ when $2M/a \gg 1$. Since $g^{zz} = g^{zz}(r)$ and $dr/dz = z/r$, the apparent horizon at the endpoint of evaporation has zero surface gravity.

The mass function $m = r(1 - z^2 g^{zz}/r^2)/2$ can be inserted in the first two of Eqs. (3.1), with the result for $T_v^v$

$$-8\pi T_v^v = -8\pi \left(T_v^v\right)_H - \frac{a^2}{r^4} + \frac{\left[3\gamma(2M)^2 + 2(1+\beta)(2M)r - \alpha r^2\right]}{\left[r^3 + \beta r a^2 + \gamma(2M)a^2\right]^2} a^2 \qquad (3.23)$$

$$+ \frac{\left[\alpha(3+\beta)r^2 + 2\alpha\gamma(2M)r - \gamma(2M)^2 + \alpha\beta a^2\right]}{\left[r^3 + \beta r a^2 + \gamma(2M)a^2\right]^2} \frac{a^4}{r^2},$$

$$-8\pi \left(T_v^v\right)_H = 4L_H \begin{Bmatrix} (1+2M/r) \\ 2(r/2M)^3 \end{Bmatrix} \frac{z^2}{r^2} \frac{\left[r^4 + \beta r^2 a^2 - \alpha \gamma a^4\right]}{\left[r^3 + \beta r a^2 + \gamma(2M)a^2\right]^2}. \qquad (3.24)$$

Also,

$$4\pi T_v^z = 4\pi \left(\frac{r}{z}\right) T_v^r = -L_H \frac{z}{r} \left(\frac{\partial t'}{\partial v}\right)_r \frac{\left[r^4 + \beta r^2 a^2 - \alpha \gamma a^4\right]}{\left[r^3 + \beta r a^2 + \gamma(2M)a^2\right]^2}. \qquad (3.25)$$

The vanishing of $T_v^z$ at $z = 0$ allows a smooth transition from inflow of (negative) energy in the black hole to outflow at constant $u$ in the white hole.

The $R_z^v = 8\pi T_z^v$ Einstein equation gives in the black hole

$$4\pi e^{\bar{\psi}_v} T_z^v = \frac{a^2}{r^4} - \frac{z}{r^2}\left(\frac{\partial \bar{\psi}_v}{\partial z}\right)_v = \frac{a^2}{r^4} -$$

$$\frac{z^2}{r^2} \begin{Bmatrix} 4L_H/r^2 \\ 4L_H r^2/(2M)^4 \end{Bmatrix} + \frac{a^2\left[\delta/(2Mr^3) + 2\varepsilon/r^3 + 3\phi(2M/r^5)\right]}{1 + a^2\left[\delta/(2Mr) + \varepsilon/r^2 + \phi(2M/r^3)\right]}. \qquad (3.26)$$

Then $T_z^z$ can be found from the identity

$$T_r^r = T_z^z = -e^{\bar{\psi}_v} T^{zv} = T_v^v - g^{zz} e^{\bar{\psi}_v} T_z^v. \qquad (3.27)$$

After the transition to the white hole, $z$ is positive and derivatives of $M$ are evaluated using Eq. (3.20) instead of Eqs. (3.16). The expression for $T_u^u$ is the same as Eq. (3.23), except that $\left(T_u^u\right)_H \equiv 0$. The retarded coordinate version of Eq. (3.25) has the opposite overall sign, so $T_u^z$ like $T_v^z$ is positive. The flow of energy through the white hole is completely contained in $T_u^r$. The white hole and black hole $g^{zz}$ are identical functions of $r$, and $e^{-\bar{\psi}_u}$ differs only in that the first factor in the black hole $e^{-\bar{\psi}_v}$ is absent in $e^{-\bar{\psi}_u}$ at all $r$. The continuity of $T_z^z$ is $C^1$ at $z = 0$, since $\left(T_v^v\right)_H$ and the contribution of the flux term to $e^{\bar{\psi}_v} T_z^v$ vanish as $z^2$ approaching $z = 0$ from the black hole.

To further clarify the black hole to white hole transition, project $T_\alpha^\beta$ onto an orthonormal tetrad with future-directed 4-velocity $u^\alpha$ and radial unit vector $n^\alpha$ pointing away from the shell. Where $g^{zz} < 0$ inside the black hole and white hole apparent horizons, and particularly in the vicinity of $z = 0$, it is natural to set $u_v = 0$, so the 4-velocity is orthogonal to a spacelike displacement at constant $z$. Since $u^z > 0$, the remaining components are

$$u^v = e^{-\bar{\psi}_v} / \sqrt{-g^{zz}}, \quad u^z = \sqrt{-g^{zz}}, \quad u_z = -1/\sqrt{-g^{zz}}. \tag{3.28}$$

The radial basis vector has $n^v > 0$ so

$$n^v = e^{-\bar{\psi}_v} / \sqrt{-g^{zz}}, \quad n^z = 0, \quad n_v = e^{\bar{\psi}_v} \sqrt{-g^{zz}}, \quad n_z = -1/\sqrt{-g^{zz}}. \tag{3.29}$$

The energy density $E$, the energy flux $F$, and the radial stress $P_r$ are

$$E = -T_z^z - \left(-g^{zz}\right)^{-1} e^{-\bar{\psi}_v} T_v^z = -T_z^z - F, \quad P_r = T_v^v - F. \tag{3.30}$$

In retarded coordinates inside the white hole apparent horizon,

$$u^u = e^{-\bar{\psi}_u} / \sqrt{-g^{zz}}, \quad u^z = \sqrt{-g^{zz}}, \quad u_z = -1/\sqrt{-g^{zz}}, \tag{3.31}$$

$$n^u = -e^{-\bar{\psi}_u} / \sqrt{-g^{zz}}, \quad n^z = 0, \quad n_u = -e^{\bar{\psi}_u} \sqrt{-g^{zz}}, \quad n_z = -1/\sqrt{-g^{zz}}. \tag{3.32}$$

The energy density, energy flux, and radial stress are

$$E = -T_z^z - \left(-g^{zz}\right)^{-1} e^{-\bar{\psi}_u} T_u^z = -T_z^z + F, \quad P_r = T_u^u + F. \tag{3.33}$$

Since $e^{-\bar{\psi}_u} T_u^z$ and $e^{-\bar{\psi}_v} T_v^z$ are identical functions of $|z|$ and the black hole and white hole frames are identical at $z = 0$, the energy flux goes smoothly $(C^1)$ from positive in the black hole to negative in the white hole.

The energy flux is singular at $g^{zz} = 0$, because the $u_v = 0$ frame is infinitely boosted relative to any local inertial frame. A simple choice of frame valid where $g^{zz} > 0$ is the static frame, defined by $u^z = 0$. Then outside the black hole

$$E = -T_v^v - \left(g^{zz}\right)^{-1} e^{-\bar{\psi}_v} T_v^z = -T_v^v - F, \quad P_r = T_z^z - F. \tag{3.34}$$

Outside the white hole

$$E = -T_u^u - \left(g^{zz}\right)^{-1} e^{-\bar{\psi}_u} T_u^z = -T_u^u + F, \quad P_r = T_z^z + F. \tag{3.35}$$

Near the apparent horizons the divergence of $F$ in the static frame implies $E \cong P_r \cong -F$ for the black hole, an inflow of negative energy, and $E \cong P_r \cong +F$ for the white hole, an outflow of negative energy. Everything is regular in a free-fall frame.

The $G_\theta^\theta$ component of the Einstein tensor is rather complicated, and $T_\theta^\theta = T_\varphi^\varphi$ can most easily be found from the $T_{r;\mu}^\mu = 0$ conservation equation. In advanced coordinates for the black hole,

$$2T_\theta^\theta = \frac{1}{r}\left(r^2 T_z^z\right) + re^{-\bar\psi_v}\left(e^{\bar\psi_v}T_r^v\right)_{,v} - \left(r\bar\psi_{v,r}g^{zz} + \frac{r}{2}g^{zz}_{,r}\right)e^{\bar\psi_v}T_z^v, \quad (3.36)$$

and similarly for the white hole. There is a small discontinuity proportional to $L_H$ in the radial derivative of $T_z^z$ at $r = 2M$ for the black hole, and therefore in $T_\theta^\theta$. $T_\theta^\theta$ is finite at $z = 0$, in spite of a singular term in $T_r^v = (r/z)T_z^v$, because the singular term does not depend on $v$, leaving only a small discontinuity from a small term in $\bar\psi_{v,r}$ without a counterpart in $\bar\psi_{u,r}$. All these discontinuities can presumably be avoided by making a more elaborate ansatz for $t'(v,r)$. The Einstein tensor at $z = 0$ in the limit $2M/a \gg 1$ is simply

$$G_v^v = (1 - 2/\gamma)a^{-2}, \; G_z^z = -a^{-2}, \; G_z^v = 2a^{-2}, \; G_v^z = 0, \; G_\theta^\theta = (4 - 5/\gamma)a^{-2}. \quad (3.37)$$

The energy density is positive, given $0 < \gamma < 1$, but the dominant energy condition is violated.

At $r \gg (2Ma^2)^{1/3}$ with $2M/a \gg 1$ there is a semi-classical regime where quantum corrections to the geometry are small and the effective stress-energy tensor is dominated by terms first-order in $\hbar$, i.e., first-order in an expansion in powers of $a^2$. In this limit the expressions for the components of the SCSET become polynomials in $x \equiv 2M/r$. Outside the black hole apparent horizon, with $L_H = q(a/2M)^2$,

$$-8\pi T_v^v = \frac{a^2}{(2M)^4}\left[4qx^2(1+x) - (1+\alpha)x^4 + 2(1+\beta)x^5 + 3\gamma x^6\right], \quad (3.38)$$

$$8\pi T_v^z = -2q\frac{a^2}{(2M)^4}x^2, \; 8\pi T_z^v = -\frac{a^2}{(2M)^4}\left[8qx^2 + 2(2\varepsilon - 1)x^3 + 6\phi x^3\right], \quad (3.39)$$

$$8\pi T_z^z = \frac{a^2}{(2M)^4}\left[\begin{array}{l}4qx^2(1-3x) + 2\delta x^3 + (\alpha - 2\delta + 4\varepsilon - 1)x^4 \\ -2(\beta + 2\varepsilon - 3\phi)x^5 - (3\gamma + 6\phi)x^6\end{array}\right], \quad (3.40)$$

$$8\pi T_\theta^\theta = \frac{a^2}{(2M)^4}\left[\begin{array}{l}(8q - \delta)x^3 - \left(\alpha - \frac{5}{2}\delta + 4\varepsilon - 1\right)x^4 \\ + \left(3\beta + 7\varepsilon - \frac{1}{2} - 9\phi\right)x^5 + \left(6\gamma + \frac{27}{2}\phi\right)x^6\end{array}\right]. \quad (3.41)$$

The trace of the stress-energy tensor is independent of $q$,

$$8\pi T_\mu^\mu \cong \frac{a^2}{(2M)^4}\left[(3\delta + 2 - 4\varepsilon)x^4 + (2\beta - 3 + 10\varepsilon - 12\phi)x^5 + (6\gamma + 21\phi)x^6\right]. \quad (3.42)$$

Numerical calculations of the Unruh state SCSET in the exterior of the black hole have been carried out for massless, conformally-coupled scalar and vector fields[28] and massless minimally coupled scalar fields[29]. These can be fit[10] within their numerical accuracy by 6th order polynomials in $2M/r$. However, they all have negative coefficients for the $x^6$ term in the energy density, corresponding to $\gamma < 0$ in my model. However, the unknown contribution to the SCSET from quantum fluctuations of the gravitational field should dominate. The Hawking luminosity for a single scalar field corresponds to $q = 2.98 \times 10^{-4}$ if $a = m_p$, and is smaller for higher-spin fields.

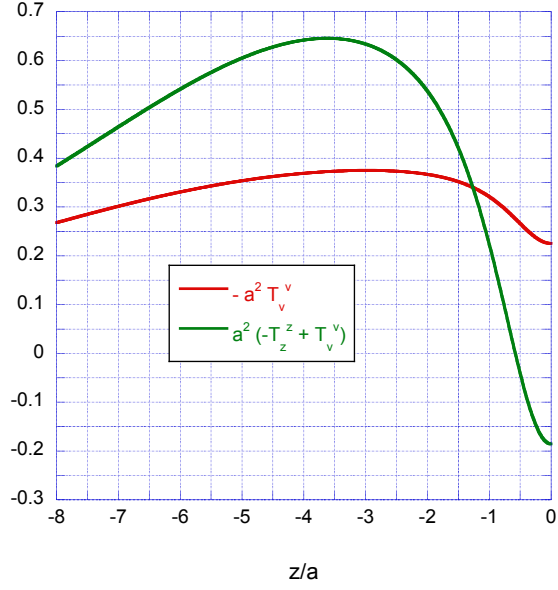

Figure 2. The components $-T_z^z$ and $-T_v^v$ in the core of a black hole for the Set A parameters when $2M/a = 8000$. Add the curves to get the effective energy density $-T_z^z$.

All that is known about the contributions to the SCSET from quantum fluctuations in the gravitational field is the Hawking luminosity and the spin 2 trace anomaly, but that is enough to give some crude guidance to the choice of parameters in my model. Adding the spin 1 and spin 2 contributions to the trace anomaly gives

$$8\pi T_\alpha^\alpha = \frac{m_p^2}{(2M)^4} \frac{199}{30\pi} x^6 = 2.11 \frac{m_p^2}{(2M)^4} x^6. \quad (3.43)$$

To the extent that the trace is dominated by the trace anomaly, the coefficients of powers of $x$ less than 6 in the trace should be small or zero in the semi-classical regime, constraining the model coefficients in Eq. (3.42). A set of model parameters, denoted as Set A, consistent with these constraints is

$$\alpha = -0.4, \ \beta = -0.7, \ \gamma = 0.3, \ \delta = 0.08, \ \varepsilon = 0.56, \ \phi = 0.1. \quad (3.44)$$

These parameters give $2M_{min}/a = 1.0$ and a coefficient of $x^6$ in the trace about twice that of Eq. (3.43) if $a = m_p$. Note that the equation for locating apparent horizons $(g^{zz} = 0)$ can be written

$$\left(\frac{r}{a}\right)^3 - \left(\frac{2M}{a}\right)\left(\frac{r}{a}\right)^2 + \gamma\left[\left(\frac{2M}{a}\right) - \left(\frac{2M}{a}\right)_{min}\left(\frac{r}{a}\right)\right] + \left[\left(\frac{2M}{a}\right)_{min} - 1\right]\left(\frac{r}{a}\right) = 0. \quad (3.45)$$

With $2M_{min}/a = 1$, the exact solution is just $r/a = 2M/a$.

Fig. 2 shows how $-T_v^v$ and $-T_z^z$ vary with $z$ at constant advanced time $v$ in the black hole core for the Set A parameters when the black hole is moderately massive, $2M/a = 8000$. Of course, this mass is incredibly small compared to the mass of any astrophysical black hole, and implies a Hawking temperature around the grand unification scale, exciting many more types of particles than just photons and gravitons. Since $g^{zz} < 0$ in the core and $F$ is small, $-T_z^z \cong E$ and $T_v^v \cong P_r$ (see Eq. (3.30)). The transition to the semi-classical regime, where quantum modifications to the geometry become negligible, begins around $|z| \approx 20a$. Note that $T_z^z = T_v^v$, due to $T_z^v$ vanishing, at about $z/a = -0.6$.

Once the black hole has evaporated down to close to the Planck scale, there is no semi-classical regime inside the horizon and the very notion of a quasi-classical geometry for the black hole core is suspect. There is no longer any real physical justification for my chosen form of the metric, let alone for any particular choice of parameters. Still, the model does demonstrate the *possibility* of an evolution in which the black hole ends and the white hole begins without any singularity.

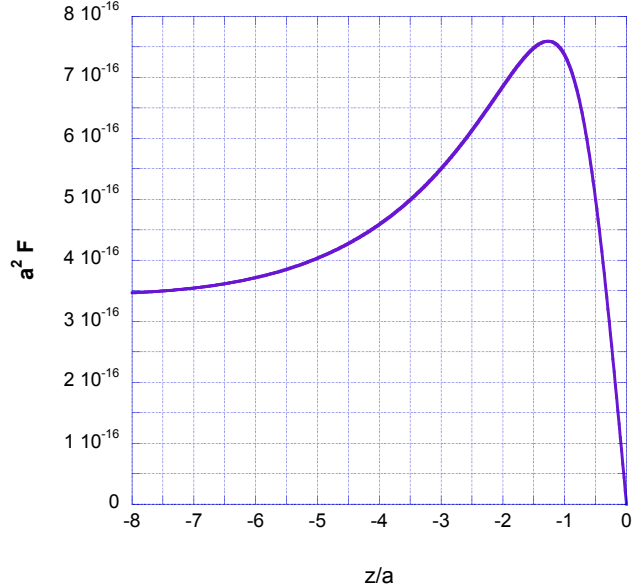

Figure 3. The energy flux in local orthonormal frames in the core of a black hole with the Set A parameters and $2M/a = 8000$, as in Fig. 2.

Rather arbitrarily setting $q = 0.0015$, about 100 times the value for photons and gravitons, the energy flux in the core of the black hole for same parameters as in Fig. 2 is plotted in Fig. 3. The energy flux is smaller here than the dominant terms in the stress-energy tensor by a factor of order $(a/2M)^4$. While in the vicinity of the black hole horizon the energy flux is not enormously smaller than the other components of the stress-energy tensor, it increases much more slowly going in toward the core.

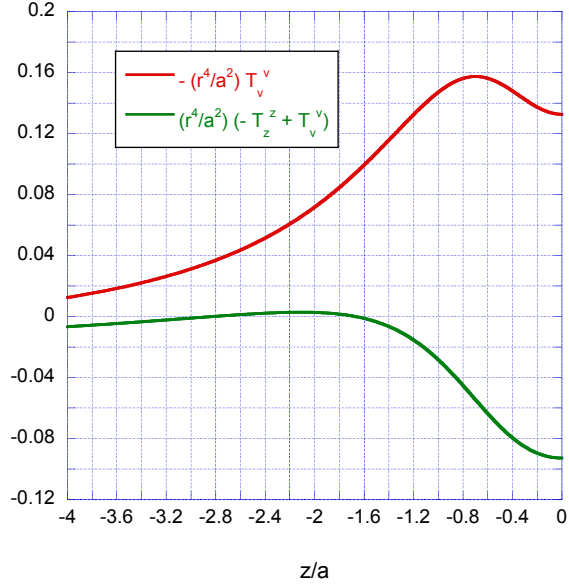

Figure 4. The dominant components of the stress-energy tensor for Set A parameters when $2M/a = 3.0$. The black hole trapping horizon at $z/a = -2.83$, .

For what it is worth, I plot in Fig. 4 the $T_v^v$ and $T_z^z$ components of the model stress-energy tensor in the core of the black hole for the Set A parameters when $2M/a = 3$. The scaling by $r^4/a^2$ is to compensate for the rapid falloff in the stress-energy tensor components once $r/a > (2M/a)^{1/3}$. The black hole trapping horizon is at $r/a = 3$, $z/a \cong -2.83$. The surface gravity of the trapping horizon at this point is just a bit smaller than the classical value of $1/4M$. $T_z^z$ also equals $T_v^v$ at $|z/a| \cong 1.67$, because $T_z^v = 0$ there.

It is possible to modify the model parameters so that the asymptotic static energy density from Eq. (3.38) is positive at all radii, while maintaining the other desirable features of the model, but the model parameters are more tightly constrained. For instance, with the parameters

$$\alpha = -1.2, \ \beta = -0.9, \ \gamma = 0.2, \tag{3.46}$$

$(2M/a)_{\min} = 1.625$ and $\partial g^{zz}/\partial r$ is just slightly positive at $r = a$.

## IV. QUANTUM ENERGY CONDITIONS

An interesting question to ask of the model is whether it is consistent with quantum energy conditions that have been proven in some generality in a semi-classical contest. One such condition is the achronal averaged null energy condition[30] (ANEC). This states that

$$\int_{-\infty}^{\infty} T_{\alpha\beta} k^\alpha k^\beta \, d\lambda \geq 0, \tag{4.1}$$

where the integral is over a complete *achronal* (no two points connected by a timelike curve) null geodesic with affine parameter $\lambda$ and tangent vector $k^\alpha = dx^\alpha / d\lambda$. I first consider radial null geodesics crossing the black hole and white hole horizons and then the null generators of the black hole and white hole horizons.

In the black hole region an "ingoing" radial null geodesic has $k^v = 0 \Rightarrow k_z = 0$ and $k^z = -e^{-\bar{\psi}_v} k_v > 0$, so

$$T_{\alpha\beta} k^\alpha k^\beta = -e^{-\bar{\psi}_v} T_z^{\ v} (k_v)^2. \tag{4.2}$$

This is non-negative except for a small Planck-scale region near $z = 0$ if $\delta, \phi \geq 0$ and $\varepsilon > 1/2$. The evaporation time scale is much longer than a dynamical time scale, and to a good approximation $k_v < 0$ is constant along the trajectory. The integral as the geodesic goes from $z = -\infty$ to $z = 0$ is

$$\int_{-\infty}^0 T_{\alpha\beta} k^\alpha k^\beta (k^z)^{-1} dz = +\int_{-\infty}^0 e^{-\bar{\psi}_v} \left( e^{\bar{\psi}_v} T_z^{\ v} \right) k_v \, dz. \tag{4.3}$$

While $e^{\bar{\psi}_v} T_z^{\ v}$ varies relatively slowly, and is negative in the core of the black hole except for $|z/a| < O(1)$, $e^{-\bar{\psi}_v}$ falls off rapidly, and the net result for the integral of Eq. (4.3) is negative.

As long as $\dot{M}(u) \geq 0$, the continuation of the "ingoing" null geodesic into the white hole region stays inside the anti-trapping horizon where $g^{zz} = 0$. An equation for the tangent vector in the retarded Eddington-Finkelstein coordinates is

$$dk^u / dz = \left( \bar{\psi}_{u,r} + g^{zz}_{\ ,r} / 2 \right) e^{\bar{\psi}_u} k^u. \tag{4.4}$$

For large $2M/a$, $k^u$ grows exponentially with a time constant $\cong 4M$. The integral of the null energy can be written as

$$\int \left[ -(dz/du)^2 T_z^{\ u} - T_u^{\ z} \right] e^{\bar{\psi}_u} k^u \, du. \tag{4.5}$$

During the growth of the white hole, corresponding to the evaporation of the black hole, $\dot{M}(u) > 0$ and $T_u^{\ z} > 0$. The first term is positive except close to $r = a$, but is suppressed near the white hole horizon as $dz/du$ becomes very small. The second term in Eq. (4.5) is negative, and while initially small compared to the first term, it quickly becomes dominant. The second term does become positive when the matter and radiation that collapsed to form the black hole starts escaping from the white hole and $\dot{M}(u) < 0$. The exponential growth of $k^u$ means that this positive contribution will dominate, since the all but the last e-folding of the negative contributions are suppressed in comparison. The same reasoning applies in time reverse applies to radially "outgoing" null geodesics passing through the collapsing matter, the interior of the black hole, and exiting across the white hole horizon.

The ANEC is also satisfied for the null generators of the black hole and white hole horizons, since the integrals are also dominated by the positive contributions as they pass through the collapsing radiation as the black hole forms and through the expanding radiation as the white hole disappears.

The quantum null energy condition[31] (QNEC) is a quasi-local lower limit on the null energy based on the von Neumann entropy $S_{vN}$ of the region outside a zero-expansion null hypersurface,

$$\langle T_{\alpha\beta} k^\alpha k^\beta \rangle \geq \frac{\hbar}{2\pi A} \frac{d^2 S_{vN}}{d\lambda^2}, \tag{4.6}$$

where $A$ is the area of a cross-section. In the present context, this can be applied at the black hole and white hole horizons, with the cross-section a two-surface of constant $r$ and, respectively, of constant $v$ or $u$. While neither horizon is exactly zero-expansion, they are close enough, at least while $2M/a \gg 1$. Then $S_{vN}(v)$ is slowly increasing on the black hole horizon and on the white hole horizon $dS_{vN}(u \cong -v)/du \cong -dS_{vN}/dv$. To a good approximation when $M \gg a$, and with surface gravity $\kappa = 1/[4M(v)]$ on the black hole horizon,

$$\frac{d^2 S_{vN}}{d\lambda^2} = \frac{d}{d\lambda}\left(\frac{dS_{vN}}{dv} k^v\right) \cong -\kappa \frac{dS_{vN}}{dv} (k^v)^2, \tag{4.7}$$

since $k^v \propto e^{-\kappa v}$ and

$$\left|(d^2 S_{vN}/dv^2)/(dS_{vN}/dv)\right| \ll -(dk^v/dv)/k^v \cong \kappa. \tag{4.8}$$

On the white hole horizon at $u \cong -v$, $k^u \cong e^{\kappa u}$ and

$$\frac{d^2 S_{vN}}{d\lambda^2} \cong \frac{d}{d\lambda}\left(\frac{dS_{vN}}{du} k^u\right) \cong +\kappa \frac{dS_{vN}}{du}(k^u)^2, \quad \kappa \frac{dS_{vN}}{du} \cong -\kappa \frac{dS_{vN}}{dv}. \tag{4.9}$$

On both horizons as long as $2M/a \gg 1$,

$$\langle T_{\alpha\beta} k^\alpha k^\beta \rangle = -\frac{L_H}{4\pi r^2}. \tag{4.10}$$

Using the Hawking luminosity and $dS_{vN}/dv$ as calculated in a semi-classical approximation by Page[32] for photons and gravitons, one can confirm that the QNEC is satisfied while the semi-classical approximation is valid, consistent with the recent claim of a quite general proof of the QNEC in a semi-classical context by Ceyhan and Faulkner[33].

The most serious potential flaw in my model would seem to be the negative energy propagating out to future null infinity from the white hole. Can the emission of the negative energy be avoided by assuming it propagates along "ingoing" null geodesics inside the white hole? This scenario is not viable. The problem is that then $M$ decreases along "outgoing" radial null geodesics. The equation for an outgoing null geodesic trajectory in advanced Eddington-Finkelstein coordinates is

$$\left(\frac{dz}{dv}\right)_u = -\frac{1}{2}e^{\bar{\psi}_u}g^{zz}. \tag{4.11}$$

Starting from $z > 0$, the decreasing $M$ causes $g^{zz}$ to change sign from negative to positive at a finite $v$. Outside this apparent horizon $z$ decreases and asymptotically approaches zero as $v \to \infty$. What this signifies is a traversable wormhole in the white hole exterior, as discussed by Simpson, et al[34] for their Fig. 4, but with the throat pushed off to future null infinity. In a conventional Schwarzschild white hole, with constant $M$, the advanced time $v \to \infty$ at the $r = 2M$ Cauchy horizon on an outgoing radial null geodesic. In the exterior, $r$ continues to increase, with $v$ now decreasing, consistent with Eq. (4.11).

The asymptotic geometry is Minkowski in my model, and for massless quantum fields in Minkowski spacetime Ford and Roman[35] have argued that a lower bound to energy density measured by an inertial observer averaged over a time $t_0$ is $E_{\min} \sim -m_p^2/t_0^4$. At a radius $r$ from a mass $M$ the time over which tidal accelerations can be neglected means $t_0$ can be as large as $r^{3/2}/M^{1/2}$, implying a minimum averaged energy density $E_{\min} \sim -\hbar M^2/r^6$. At $r \gg M$ this bound is substantially violated by the negative energy density associated with the negative energy flux from the white hole in my model, of order $-\hbar/(M^2 r^2)$. Some reasons why the Ford-Roman bound might not apply in the black hole to white hole scenario are discussed in Part V.

Bianchi and Smerlak[36] have made arguments, based on a 2D approximation to black hole evaporation, that an episode of negative energy outflow to future null infinity is *required* in any unitary black hole evaporation scenario. Their result is a necessary condition for unitary evolution of the black hole, in which the von Neumann entropy of the exterior is initially and finally zero,

$$\int_{-\infty}^{\infty} \dot{M}(u)\exp[6S_{vN}(u)]du = 0. \tag{4.12}$$

This condition is trivially satisfied for my model, but it can also be satisfied by a brief episode of emission of negative energy when the entropy is near its maximum that would not necessarily violate the Ford-Roman bound.

Finally, the exponentially increasing blueshift of any external energy propagating along the white hole horizon should not be a problem. There is no reason for a substantial amount of such energy in the context of my model, since the only source for an isolated white hole is the backscatter off of the background curvature of the outgoing Hawking radiation from the black hole and of the outgoing negative energy radiation from the white hole. The stress-energy tensor of a null fluid is $T^{\alpha\beta} = \sigma k^\alpha k^\beta$, where $k^\alpha$ is a null tangent vector obeying the geodesic equation. In the retarded coordinates when the geometry is close to Schwarzschild the geodesic equation gives $dk^u/du \cong (M/r^2)k^u \cong \kappa k^u$ close to the horizon, with the solution $k^u \cong (k^u)_0 e^{\kappa u}$. Then $k^r = -(1-2M/r)k^u/2$, from which $r - 2M \cong (r - 2M)_0 e^{-\kappa u}$, $k_u \cong -\kappa (r - 2M)_0 (k^u)_0$ and $k_r \cong -k^u$. Conservation of the stress-energy gives $d\sigma/du + \sigma k^\alpha_{;\alpha}/k^u = 0$. Since $k^\alpha_{;\alpha} = (2/r)(dr/du)k^u$, $d\sigma/du = (r - 2M)\sigma/r^2 \propto e^{-\kappa u}$ and $\sigma \to \sigma_0$, a constant. The contribution to the mass function $m$ from the stress-energy tensor on the horizon is

$$\Delta m \sim -16\pi M^2 \sigma_0 \int k^u k_u \, dr \sim +2\pi M \sigma_0 (r - 2M)^2 (k^u)^2, \qquad (4.13)$$

which is constant in spite of the exponential blueshift, as is required by energy conservation. The change in $e^{-\psi_u}$ across the horizon is also unaffected by the blueshift. Of course, these are classical estimates that do not preclude quantum instabilities. However, the blueshift is locally just an artifact of evaluating the energy in frames accelerating in the opposite direction from the direction of the flow of energy along the horizon. To the extent that the quantum theory is invariant under local Lorentz transformations, such quantum instabilities should not be present.

The concern expressed in Ref. [25] that energy propagating along the white hole horizon would cause conversion of the white hole into a black hole when it intersects the outgoing shell of rebounding radiation (at $u = u_2$ in Fig 1) is not an issue, since at that point the backscatter should be predominantly of negative energy emitted from the white hole, which works against geodesic convergence.

## V. DISCUSSION

At best the toy model I have constructed is only representative of the many quasi-classical histories contributing to the wave function of the black hole. A full quantum gravity treatment is required for any final resolution of the fate of a black hole and the information problem. While my model seems consistent with the existing framework for LQG calculations, it does require quite different values for some of the LQG parameters than those usually adopted in the literature. With my choice of parameters, the minimum two-sphere area in the black hole interior is a Planck scale constant related directly to the fundamental "area gap" parameter of LQG and is independent of the mass of the black hole. While quantum tunneling to the white hole from a radius large compared to the Planck scale might be possible, I would expect the quantum amplitude would be very small compared to that of nonsingular quasi-classical evolution.

I argue that it is reasonable to consider the quantum geometry as small fluctuations about a quasi-classical geometry as long as $r \gg a$, even if this background geometry is substantially modified from a classical solution of the vacuum Einstein equations by quantum backreaction. The effective stress-energy tensor in this quasi-classical geometry is derived from the Einstein tensor calculated from the model metric tensor and is considered to include the macroscopic effects of quantum fluctuations in the gravitational field as well as those of non-gravitational fields. This can make sense as long as individual modes of the quantum fields are small perturbations of a background geometry, even though the cumulative effect of a large number of these modes may substantially modify the geometry. In the context of Schwarzschild, the semi-classical approximation of a fixed classical background geometry should be valid where the spacetime curvature is very sub-Planckian, $M/r^3 \ll m_p^{-2}$, or $r \gg \left(Mm_p^2\right)^{1/3}$.

While my guess at the form of the metric in the quasi-classical regime is rather ad hoc, when extrapolated beyond the black hole horizon it can match the general form of the SCSET found by numerical calculations in the literature for spin 0 and spin 1 fields in the Unruh state, but not necessarily the precise values of all the coefficients. The geometry in the model varies smoothly in the transition between the black hole and the white hole throughout the black hole evaporation, even when the black hole horizon area is close to the Planck scale. Of course, one expects large quantum fluctuations in the geometry where $r/a$ is of order one. It would not be surprising if the QNEC were violated there, since it is basically a semi-classical result. Quantum singularity theorems, such as the quantum focusing conjecture[37], (QFC) also presumably would not apply in a highly quantum regime.

The most disturbing feature of my model is that the white hole evolves for most of its lifetime by emitting negative energy. This is the same negative energy that flows inward across the black hole horizon during evaporation, as indicated by the calculations of the SCSET. I argued in Part IV that retaining the negative energy inside the white hole is inconsistent with a reasonable asymptotically flat geometry outside the white hole horizon. Also, without emitting negative energy the Planck scale white hole formed from a black hole that evaporates down to the Planck scale must remain Planck scale, which means that somehow all the rebounding matter and radiation that formed the black hole must emerge from the white hole with a net Planck scale energy. I do not see how a physically sensible black hole to white hole scenario is possible without emission of negative energy from the white hole.

Is there some way to rationalize the gradual outflow of negative energy from the white hole? The generation of Hawking radiation should be thought of as the tidal disruption of vacuum fluctuations in the vicinity of the black hole horizon, part of which propagate to future null infinity directly with positive energy and part of which end up inside the black hole with negative energy. These parts are not independent of each other. They are strongly entangled and correlated. If the part inside the black hole later propagates out of the white hole to future null infinity, it does not do so as normal "particles", which must have positive energy in the asymptotic Minkowski region. The negative energy emissions together with the earlier Hawking radiation are still parts of vacuum fluctuations, albeit *very* highly distorted by the black hole.

A somewhat similar situation arises for a zero-energy vacuum fluctuation straddling and propagating along a null hypersurface in Minkowski spacetime. A uniformly accelerating observer for whom that hypersurface is a Rindler horizon becomes infinitesimally close to the horizon in the original inertial frame and only part of the fluctuation is accessible to him. If he eventually stops accelerating, he will gain access to the hidden part of the fluctuation, and be able to verify that the energy of the entire fluctuation is zero, but until then the part he can observe may have a small non-zero energy. Important differences from the black hole horizon are no systematic preference in the sign of the energy averaged over many such fluctuations and no conflict the Ford-Roman bound. The Unruh thermal radiation measured by an accelerating particle detector is not relevant here, since this is a property of the *detector* interacting with the vacuum, and has nothing to do with the stress-energy tensor that is the source in the Einstein equations.

My scenario is incomplete, since there is no explicit modeling of how the bounce proceeds in the interior of the star or shell that collapses to form the black hole. There must be some sort of transition from a minimum radius of zero on the timelike trajectory of the very center of a star or shell that collapses to form the black hole to a $r = a$ minimum radius outside the star or shell. What is depicted in Fig. 1 is nothing more than a crude and very schematic guess.

If the black hole does evaporate down to the Planck scale, with no significant release of quantum information across the black hole horizon, as I assume, it is apparent that the Bekenstein-Hawking entropy[5] $S_{BH} = A/(4\hbar) = 4\pi(M/m_p)^2$ should *not* be interpreted as a measure of the total number of quantum degrees of freedom associated with the black hole. The "partners" of the Hawking radiation quanta simply cross from the black hole region to the white hole region in Fig. 1 and then flow outward across the white hole horizon. Near the end of the black hole evaporation $S_{BH}$ is tiny compared with the entropy of the Hawking radiation and the von Neumann entropy of the black hole exterior. It is a mistake to think of the black hole interior degrees of freedom as being in any kind of thermal equilibrium. The degrees of freedom of the bouncing shell and entangled vacuum modes crossing the $z = 0$ spacelike hypersurface are completely out of causal contact with the horizon degrees of freedom of the late stages of the black hole evaporation. While $S_{BH}$ is presumably a measure of the maximum number of quantum degrees of freedom that are present near and on the black hole horizon at any one time, quantum fluctuations on the horizon do not stay on the horizon. They end up partially in the Hawking radiation and partially deep inside the black hole. Similar views have been recently expressed by Garfinkle[38] and by Rovelli[39].

Finally, the assumption of spherical symmetry is unrealistic. Any small deviations from spherical symmetry in the collapse that forms the black hole are amplified as the collapse proceeds, and classically the singularity structure of a Kerr black hole with any nonzero angular momentum is timelike, rather than the spacelike singularity of a Schwarzschild black hole. So does the black hole to white hole transition discussed here have any relevance to an even slightly generic black holes? Bianchi and Haggard[40] have made an initial attempt to address this question. They argue that at least the initial breakdown of the semi-classical approximation in black holes and the onset of the quantum gravitational regime for quantum geometries with realistic admixtures of zero and nonzero angular momentum is always spacelike in character.

ACKNOLEDGEMENTS

This paper was inspired by discussions with Hal Haggard while we were both visiting the Perimeter Institute. Research at the Perimeter Institute is supported by the Government of Canada through the Department Innovation, Science, and Economic Development, and by the Province of Ontario through the Ministry of Research and Innovation. I also thank Amos Ori and Tommaso De Lorenzo for comments on an earlier version.